\documentclass[aps,prl,reprint,groupedaddress,showpacs]{revtex4-1}
\usepackage{graphicx}
\usepackage{textcomp}
\usepackage{hyperref}
\usepackage{ifthen} % provides \ifthenelse test  
\usepackage{xifthen} % provides \isempty test
\usepackage{amssymb,amsmath} %bold fonts in math mode

\newcommand{\fig}[2][]{%
\ifthenelse{\isempty{#1}}
{Fig.~\ref{#2}}% if no subfigure is given
{Fig.~\ref{#2}(#1)}% else
}

\begin{document}

\title{Non-adiabatic dynamics of two strongly coupled nanomechanical resonator modes}

\author{Thomas Faust}
\author{Johannes Rieger}
\author{Maximilian J. Seitner}
\author{Peter Krenn}
\author{J\"org P. Kotthaus}
\email[]{kotthaus@lmu.de}
\author{Eva M. Weig}
\email[]{weig@lmu.de}
%\homepage[]{Your web page}
%\thanks{}
%\altaffiliation{}
\affiliation{Center for NanoScience (CeNS) and Fakult\"at f\"ur Physik, Ludwig-Maximilians-Universit\"at, Geschwister-Scholl-Platz 1,
M\"unchen 80539, Germany}

\begin{abstract}
The Landau-Zener transition is a fundamental concept for dynamical quantum systems and has been studied in numerous fields of physics. Here we present a classical mechanical model system exhibiting analogous behaviour using two inversely tuneable, strongly coupled modes of the same nanomechanical beam resonator.
In the adiabatic limit, the anticrossing between the two modes is observed and the coupling strength extracted.
Sweeping an initialized mode across the coupling region allows mapping of the progression from diabatic to adiabatic transitions as a function of the sweep rate.
%[585 of 600 characters]
\end{abstract}


\maketitle

The time dynamics of two strongly coupled harmonic oscillators follows the Landau-Zener model\,\cite{L.D.Landau1932,Zener1932,Stueckelberg1932,springerlink:10.1007/BF02960953}, which is used to describe the quantum mechanical mode tunneling in a non-adiabatic transition.
This phenomenon is observed and utilized in many areas of physics, e.\,g. atomic resonances\,\cite{PhysRevA.23.3107}, quantum dots\,\cite{Petta2010}, superconducting qubits\,\cite{PhysRevLett.96.187002} and nitrogen-vacancy centers in diamond\,\cite{Fuchs2011a}.
It is also possible to create classical model systems exhibiting the same time evolution, which until now have been restricted to optical configurations\,\cite{1990PhRvL..65.2642S,1995PhRvA..51..646B}.
Such systems are well suited for the study of diabatic behaviour over a wide parameter space; for example nonlinearities could be readily introduced, potentially leading to chaotic behaviour\,\cite{1990PhRvL..65.2642S,PhysRevE.51.1861}.

Nanomechanical resonators with frequencies in the MHz range can be realized with high mechanical quality factors\,\cite{verbridge:124304,PhysRevLett.105.027205} and easily tuned\,\cite{Unterreithmeier2009} in frequency.
This makes them particularly well-suited for exploration of their coupling to other mechanical, optical or electrical microwave resonators.
Strong cavity coupling in the optical or microwave regime has been widely studied as it enables both cooling and self-oscillation of the mechanical modes\,\cite{Groblacher2009,2011arXiv1107.3761V,Teufel2011,Chan2011}.
In addition, the time-resolved Rabi oscillations between a strongly coupled two-level system and a micromechanical resonator have been observed\,\cite{OConnell2010}.

Purely mechanical, static coupling between different resonators\,\cite{APEX.2.062202,PhysRevB.79.165309,PhysRevLett.106.094102,perisanu:063110} and between different harmonic modes of the same resonator\,\cite{PhysRevLett.105.117205} has also been demonstrated.
Here, we explore the coupling between the two fundamental flexural modes\,\cite{kozinsky:253101} of a single nanomechanical beam vibrating in plane and out of plane, respectively.
We study the adiabatic to non-adiabatic transitions between the two strongly coupled classical mechanical modes in time-dependent experiments, in correspondence to the Landau-Zener transition.

\begin{figure}
\includegraphics{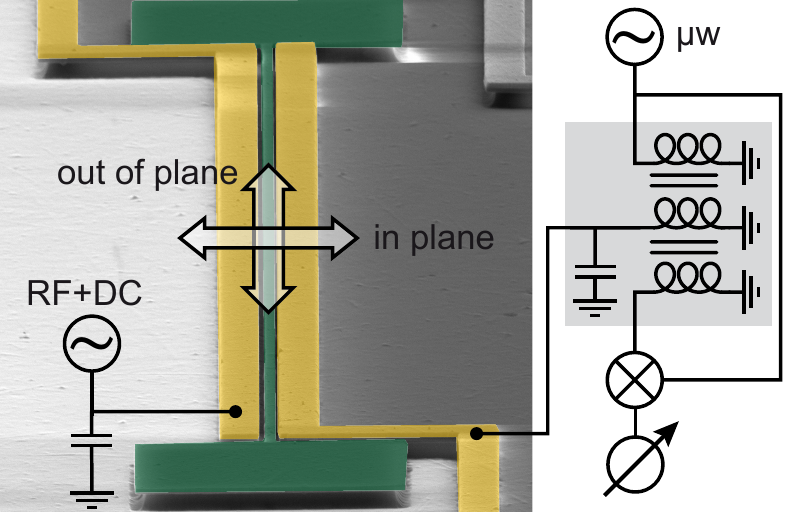}
\caption{\label{1}(color online). The SEM micrograph of the 55\,\textmu m long and 260\,nm wide silicon nitride string resonator (green) taken at an angle of $85^\circ$ also depicts the two adjacent gold electrodes (yellow) used to dielectrically drive, tune and read out the resonator motion.
The arrows denote the two mechanical modes, one oscillating parallel and the other perpendicular to the plane of the chip.
The simplified measurement scheme\,\cite{2011arXiv1109.1156F} shows the connection of the electrodes to the read-out cavity (gray box) and the microwave bypass capacitor in the bottom left.}
\end{figure}

The nanomechanical high stress silicon nitride string used in this work is shown in \fig{1}.
Two parallel gold electrodes vertically offset to the beam are used to dielectrically couple the beam oscillation to an external microwave cavity with a quality factor of $\approx 70$ at a resonance frequency of 3.44\,GHz\,\cite{2011arXiv1109.1156F}.
Displacement of the resonator leads to a change in capacitance between the two electrodes, thereby detuning the resonance frequency of the microwave circuit and creating sidebands with a frequency offset equal to the mechanical eigenfrequency.
The inductively coupled microwave cavity is driven by a signal generator; the transmission signal is demodulated and fed to a spectrum analyzer as depicted in \fig{1} and described in more detail in\,\cite{2011arXiv1109.1156F}.
In addition, a microwave bypass capacitor is used in the ground connection of one electrode which allows application of additional DC bias and RF voltages to the electrodes.
This is used to actuate the mechanical resonator via the dielectric driving mechanism\,\cite{schmid:163506,Unterreithmeier2009}.
At the same time, the dielectric coupling provides a way to tune the resonance frequency of the two mechanical modes:
The static electric field between the electrodes polarizes the dielectric resonator material which is then attracted to high electric fields, thereby changing the spring constant of the modes via the resulting force gradient\,\cite{Unterreithmeier2009}.
In the chosen geometry, where the bottom of the electrodes is flushed with the top of the beam\,\cite{2011arXiv1109.1156F}, a rising DC bias voltage causes the frequency of the in-plane mode to decrease and the out-of-plane frequency to increase\,\cite{APL}.
All experiments are performed at room temperature at pressures below $5\times10^{-4}$\,mbar.

\begin{figure}
\includegraphics{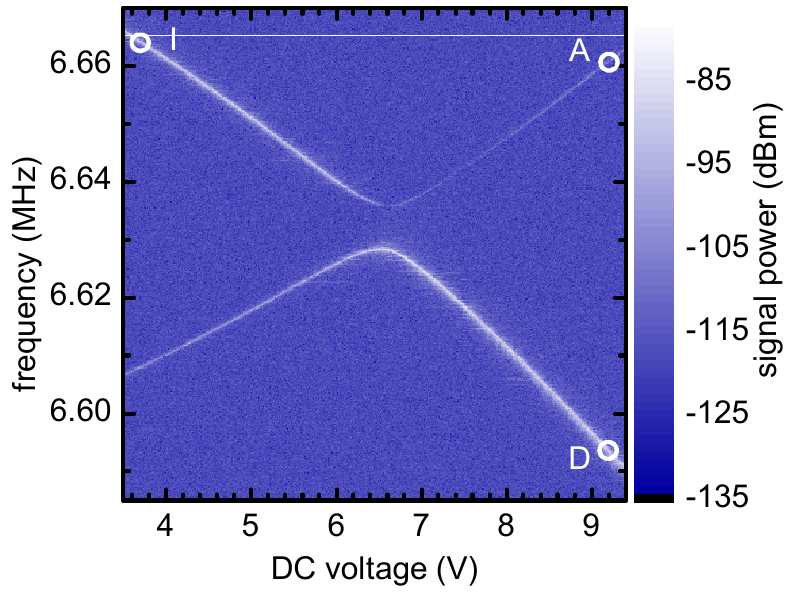}
\caption{\label{2}(color online). Both mechanical modes can be tuned in opposite direction by increasing the DC bias voltage applied to the electrodes.
The signal power of the driven resonances is shown color-coded versus DC voltage and drive frequency.
Note the clear avoided crossing between the two modes.
The three circles denote the initial state (I) and two possible final states after an adiabatic (A) or diabatic (D) transition through the coupling region, as described in the text.}
\end{figure}

At low DC bias voltages, the in-plane mode of the 55\,\textmu m long beam has a higher resonance frequency than the out-of-plane mode. This is a result of the 260\,nm beam width exceeding the beam's thickness of 100\,nm, which leads to a higher rigidity for the in-plane mode\,\cite{PhysRevLett.105.027205}.
Thus, by increasing the DC bias voltage, we are able to tune the two modes into resonance at a common frequency of approximately 6.63\,MHz.
The coupling between the modes has been observed for several resonators on various chips and is at least partially caused by the spatially inhomogenous electric field\,\mbox{\cite{supp}}. There might also be an additional, purely mechanical coupling mediated by the prestress in the beam.
The characteristic avoided crossing diagram of two coupled oscillators can be obtained by measuring the driven response of the two modes at different DC bias voltages, as shown in \fig{2}.

\begin{figure}
\includegraphics{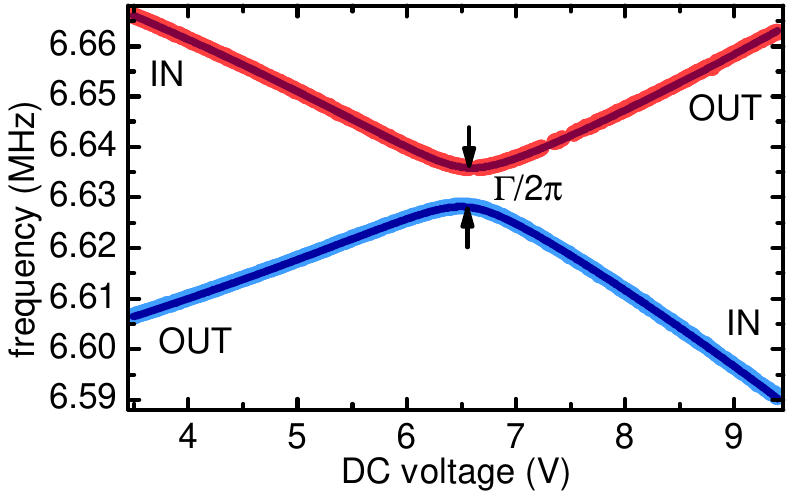}
\caption{\label{3}(color online). Frequency of the upper (red) and lower (blue) branch versus DC bias voltage. Each dot represents a value extracted from a Lorentzian fit of the data shown in \fig{2}, the solid lines are a fit of the theoretical model described in the text. IN and OUT denote the in- and out-of-plane mode of the beam.}
\end{figure}

Splitting this diagram into an upper and lower branch and fitting each dataset with a Lorentzian allows the extraction of the resonance frequencies and quality factors for each DC bias voltage applied to the electrodes.
Both modes exhibit a quality factor of approximately 80\,000, somewhat lower than in previous measurements\,\cite{2011arXiv1109.1156F}, presumably caused by fabrication imperfections.
The eigenfrequencies extracted from the anticrossing diagram are depicted in \fig{3}.
A few data points around $6.5$ and $7.4$\,V in the upper branch were omitted because of an insufficient signal to noise ratio.

For our system, the standard model of two coupled harmonic oscillators\,\cite{10.1119/1.3471177} needs to be expanded, as both oscillators react differently to the tuning parameter (the DC bias voltage).
We use the generalized differential equation for the displacement $u_n$ of each mode $n$ ($n=1,2$)
\begin{equation}
\label{deq}
m_{\rm eff}u_n''+m_{\rm eff}\gamma u_n'+k_{nm}u_n=0
\end{equation}
with 
\begin{equation}
k_{nm}=\left(
   \begin{array}{cc}
     k_1+k_c & -k_c \\
     -k_c & k_2+k_c
   \end{array}
\right),
\end{equation}
where $m_{\rm eff}$ denotes the effective mass and $\gamma=\omega/Q$ the damping constant of the resonator (identical for both modes), $k_c$ the coupling between the two modes and $k_n$ the spring constant of the respective mode.
As the DC bias voltage polarizes the resonator material and creates an electric field gradient, the additional force gradient seen by the beam depends on the square of the voltage.
We use a second-order series expansion around $U_0$ to describe the tuning behaviour:
$k_n=k_0+\kappa_n(U-U_0)+\lambda_n(U-U_0)^2$ with $\kappa_n$ and $\lambda_n$ as linear and quadratic tuning constants, assuming that both modes have the same spring constant $k_0$ at the voltage $U_0$ corresponding to zero detuning.
Note that the influence of the quadratic term is less than 15\,\% in the whole voltage range\,\cite{supp}.
The two solutions of the differential equation (\ref{deq}) describe the two branches, their fit to the experimental data is shown as solid lines in \fig{3}.
The extracted frequency splitting 
\begin{equation}
\frac{\Gamma}{2\pi}=
\frac{1}{2\pi}\left(\sqrt{\frac{k_0+2k_c}{m_{\rm eff}}}-\sqrt{\frac{k_0}{m_{\rm eff}}}\right)=
7.77\,{\rm kHz}
\end{equation}
at $U_0$=6.547\,V is much larger than the linewidth of $\gamma/2\pi=83$\,Hz, thus the system is clearly in the strong-coupling regime.

When slowly (adiabatically) tuning the system through the coupling region, the system energy will remain in the branch in which it was initialized, thereby transforming an out-of-plane oscillation to an in-plane motion (and vice-versa for the other mode).
At high tuning speeds, the diabatic behavior dominates and there is no mixing between the modes.
This classical behaviour\,\cite{10.1119/1.3471177} is analogous to the well-known quantum mechanical Landau-Zener transition.
The transition probabilities are identical in the quantum and classical case:
\begin{equation}
\label{pdia}
P_{\rm dia}=\exp\left(-\frac{\pi\Gamma^2}{2\alpha}\right),\ P_{\rm adia}=1-P_{\rm dia}
\end{equation}
where the change of the frequency difference between the two modes in time \begin{equation}
\alpha=\frac{\partial (\omega_1-\omega_2)}{\partial t}\ {\rm using}\ \omega_n=\sqrt{\frac{k_0+k_n}{m_{\rm eff}}}
\end{equation}
denotes the tuning speed\,\cite{supp}.

\begin{figure}
\includegraphics{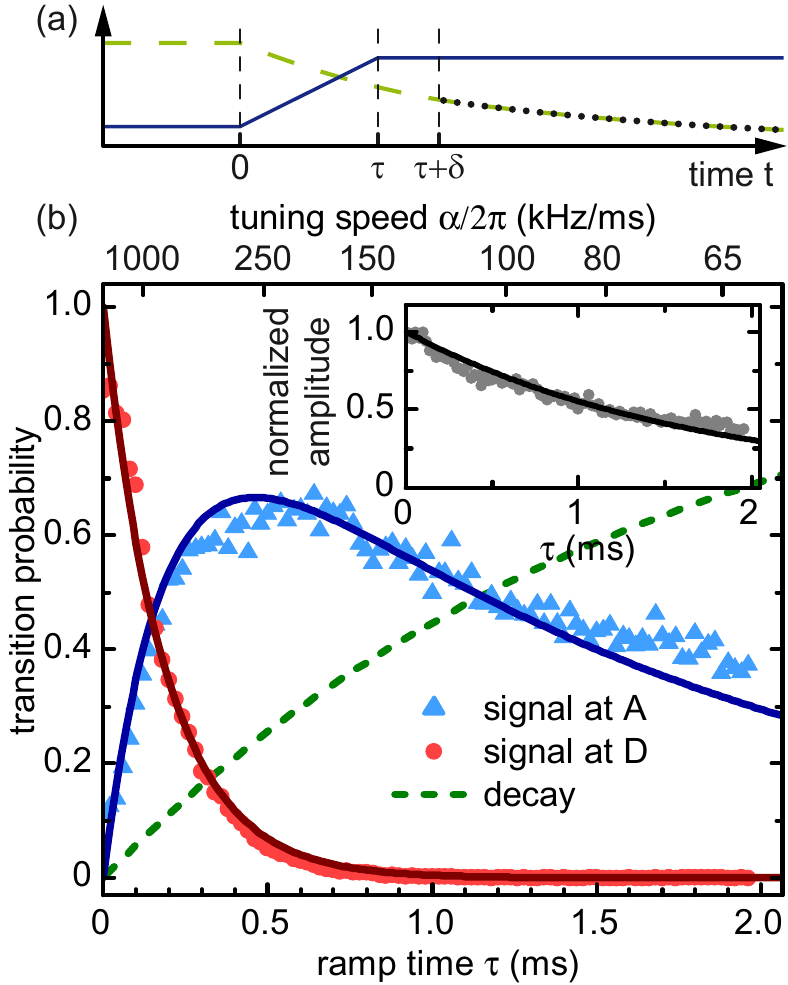}
\caption{\label{4}(color online). The measurement sequence of the time-resolved experiment is shown in (a).
At $t=0$, the DC bias voltage (blue line) is ramped up in the timespan $\tau$, after the delay $\delta$ the measurement of the mechanical signal power (green dashed line) starts at point A or D in \fig{2} and a fit (black dotted line) is used to extract the magnitude of the beam oscillation at $t=\tau$.
The normalized signal power at $t=\tau$ and thus the transition probability obtained for different ramp times $\tau$ measured at point A in \fig{3} (blue triangles) or point D (red dots) is plotted in (b) together with the theoretical model described in the text (solid lines).
The inset shows the sum of both measurements and displays a clear exponential decay.
The corresponding decay probability is represented by a green dashed line in the main plot.}
\end{figure}

The measurement sequence is depicted in \fig[a]{4}:
The system is initialized at point I (see \fig{2}) by applying a 6.6647\,MHz tone and a DC bias voltage of 3.6\,V to the electrodes.
At $t=0$, the voltage (blue line) is now ramped up to 9.1\,V within time $\tau$.
As the start and stop frequencies are kept constant throughout the experiment, changing $\tau$ changes the tuning speed $\alpha$ and therefore the transition probability.
Thus, the system's energy is distributed between point A or D (see \fig{2}), depending on the ramp time $\tau$.
At $t=0$, the mechanical resonator gets detuned from the constant drive frequency. Therefore its energy starts to decay as reflected by the decreasing signal power (green dashed line in \fig[a]{4}).
After a short additional delay of $\delta$ (to avoid transient artifacts in the measurement), the decay of the mechanical oscillation is recorded with the spectrum analyzer.
An exponential fit to the signal power, symbolized by the dotted black line in \fig[a]{4} allows the extraction of the oscillation magnitude at $t=\tau$ which is normalized to the magnitude measured before the transition at point I to account for slight variations in the initialization.
This experiment is repeated with many different ramp times $\tau$ and with the detection frequency of the spectrum analyzer set to monitor either point A or D.
The results of these measurements are shown in \fig[b]{4}.
The data clearly shows the expected behaviour:
For short ramp times below 0.2\,ms, the diabatic behaviour dominates.
For long ramp times, the adiabatic transition prevails, even though mechanical damping decreases the signal for large $\tau$.

As can be seen in the inset of \fig[b]{4}, the sum of the two curves perfectly follows the exponential decay of the mechanical energy (solid line).
This decay in amplitude between $t=0$ and $t=\tau$ has to be accounted for in the theoretical model and therefore an additional decay term $e^{-\gamma t}$ is introduced to equations\,(\ref{pdia}).
The solid lines in \fig[b]{4} show the resulting transition probabilities to point A and D and are calculated by using the $\alpha$ and $\Gamma$ obtained from the data in \fig{3}.
The measured data was rescaled by a constant factor with no free parameters to represent the probability distribution of the resonator's energy after a transition\,\cite{supp}.
A third state, representing the probability that the mechanical energy decays, is required to keep the sum of the probabilites at one.
It is determined from the inset and shown as a dashed green line in \fig{4}.
The corresponding decay constant $1/\gamma=1.92$\,ms is identical to the one extracted from the spectrally measured quality factor.
Note that dynamics with a time constant much smaller than $1/\gamma$ are observed, demonstrating coherent control of the system.

In conclusion, we utilize the strong coupling between two orthogonal modes of the same nanomechanical resonator by tuning these two modes into resonance to analyze their time-dependent dynamics.
After characterizing the coupling, we are able to model the time-resolved transition behaviour between the two modes.
The entire dynamic range between fast and coherent diabatic and slow adiabatic passages is accessible in the experiment, and a good agreement between theory and experiment is observed.

The experiment is conducted with approximately $10^9$~phonons in the vibrational mode of the resonator, thus not the single-particle probability function but the energy distribution of the ensemble is measured in the classical limit.
Since the (strongly) nonlinear regime of the utilized nanomechanical resonator can be easily accessed, the presented system could also be used to study the coupling and the time-dependent transitions of two nonlinear oscillators\,\cite{PhysRevA.66.023404,1367-2630-10-7-073008} and the development of chaotic behaviour\,\cite{PhysRevE.51.1861,1367-2630-10-7-073008} in the classical regime.
Combining cavity-pumped self-oscillation\,\cite{2011arXiv1107.3761V} with the coupled resonator modes presented here allows the study of synchronisation and collective dynamics in nanomechanical systems, as theoretically predicted\,\cite{PhysRevLett.107.043603,2011arXiv1111.2967L}.
Furthermore, after the recent breakthrough in the ground state cooling of mechanical resonators\,\cite{OConnell2010,Teufel2011,Chan2011}, the coupling between two quantum mechanical elements becomes accessible.

\begin{acknowledgments}
Financial support by the Deutsche Forschungsgemeinschaft via Project No. Ko 416/18, the German Excellence Initiative via the Nanosystems Initiative Munich (NIM) and LMUexcellent, as well as the European Commission under the FET-Open project QNEMS (233992) is gratefully acknowledged.
We thank Darren R. Southworth for critically reading the manuscript.
\end{acknowledgments}

\bibliography{mode_coupling}

%Merlin.mbs v4.21 2009-07-09.
\begin{thebibliography}{10}%
\makeatletter
\providecommand \@ifxundefined [1]{%
 \ifx #1\undefined \expandafter \@firstoftwo
 \else \expandafter \@secondoftwo
\fi
}%
\providecommand \@ifnum [1]{%
 \ifnum #1\expandafter \@firstoftwo
 \else \expandafter \@secondoftwo
\fi
}%
\providecommand \enquote [1]{``#1''}%
\providecommand \bibnamefont  [1]{#1}%
\providecommand \bibfnamefont [1]{#1}%
\providecommand \citenamefont [1]{#1}%
\providecommand\href[0]{\@sanitize\@href}%
\providecommand\@href[1]{\endgroup\@@startlink{#1}\endgroup\@@href}%
\providecommand\@@href[1]{#1\@@endlink}%
\providecommand \@sanitize [0]{\begingroup\catcode`\&12\catcode`\#12\relax}%
\@ifxundefined \pdfoutput {\@firstoftwo}{%
 \@ifnum{\z@=\pdfoutput}{\@firstoftwo}{\@secondoftwo}%
}{%
 \providecommand\@@startlink[1]{\leavevmode\special{html:<a href="#1">}}%
 \providecommand\@@endlink[0]{\special{html:</a>}}%
}{%
 \providecommand\@@startlink[1]{%
  \leavevmode
  \pdfstartlink
   attr{/Border[0 0 1 ]/H/I/C[0 1 1]}%
   user{/Subtype/Link/A<</Type/Action/S/URI/URI(#1)>>}%
  \relax
 }%
 \providecommand\@@endlink[0]{\pdfendlink}%
}%
\providecommand \url  [0]{\begingroup\@sanitize \@url }%
\providecommand \@url [1]{\endgroup\@href {#1}{\urlprefix}}%
\providecommand \urlprefix [0]{URL }%
\providecommand \Eprint[0]{\href }%
\@ifxundefined \urlstyle {%
  \providecommand \doi [1]{doi:\discretionary{}{}{}#1}%
}{%
  \providecommand \doi [0]{doi:\discretionary{}{}{}\begingroup
  \urlstyle{rm}\Url }%
}%
\providecommand \doibase [0]{http://dx.doi.org/}%
\providecommand \Doi[1]{\href{\doibase#1}}%
\providecommand \bibAnnote [3]{%
  \BibitemShut{#1}%
  \begin{quotation}\noindent
    \textsc{Key:}\ #2\\\textsc{Annotation:}\ #3%
  \end{quotation}%
}%
\providecommand \bibAnnoteFile [2]{%
  \IfFileExists{#2}{\bibAnnote {#1} {#2} {\input{#2}}}{}%
}%
\providecommand \typeout [0]{\immediate \write \m@ne }%
\providecommand \selectlanguage [0]{\@gobble}%
\providecommand \bibinfo [0]{\@secondoftwo}%
\providecommand \bibfield [0]{\@secondoftwo}%
\providecommand \translation [1]{[#1]}%
\providecommand \BibitemOpen[0]{}%
\providecommand \bibitemStop [0]{}%
\providecommand \bibitemNoStop [0]{.\EOS\space}%
\providecommand \EOS [0]{\spacefactor3000\relax}%
\providecommand \BibitemShut [1]{\csname bibitem#1\endcsname}%
%</preamble>
\bibitem{L.D.Landau1932}%
  \BibitemOpen
  \bibfield{author}{%
  \bibinfo {author} {\bibfnamefont{L.~D.}\ \bibnamefont{Landau}},\ }%
  \bibfield{journal}{%
  \bibinfo {journal} {Phys. Z. Sowjetunion}\ }%
  \textbf{\bibinfo {volume} {2}},\ \bibinfo {pages} {46} (\bibinfo {year}
  {1932})%
  \bibAnnoteFile{NoStop}{L.D.Landau1932}%
\bibitem{Zener1932}%
  \BibitemOpen
  \bibfield{author}{%
  \bibinfo {author} {\bibfnamefont{C.}~\bibnamefont{Zener}},\ }%
  \bibfield{journal}{%
  \bibinfo {journal} {Proc. R. Soc. Lond. A}\ }%
  \textbf{\bibinfo {volume} {137}},\ \bibinfo {pages} {696} (\bibinfo {year}
  {1932})%
  \bibAnnoteFile{NoStop}{Zener1932}%
\bibitem{Stueckelberg1932}%
  \BibitemOpen
  \bibfield{author}{%
  \bibinfo {author} {\bibfnamefont{E.~C.~G.}\ \bibnamefont{Stueckelberg}},\ }%
  \bibfield{journal}{%
  \bibinfo {journal} {Helvetica Physica Acta}\ }%
  \textbf{\bibinfo {volume} {5}},\ \bibinfo {pages} {369} (\bibinfo {year}
  {1932})%
  \bibAnnoteFile{NoStop}{Stueckelberg1932}%
\bibitem{springerlink:10.1007/BF02960953}%
  \BibitemOpen
  \bibfield{author}{%
  \bibinfo {author} {\bibfnamefont{E.}~\bibnamefont{Majorana}},\ }%
  \bibfield{journal}{%
  \bibinfo {journal} {Il Nuovo Cimento}\ }%
  \textbf{\bibinfo {volume} {9}},\ \bibinfo {pages} {43} (\bibinfo {year}
  {1932})%
  \bibAnnoteFile{NoStop}{springerlink:10.1007/BF02960953}%
\bibitem{PhysRevA.23.3107}%
  \BibitemOpen
  \bibfield{author}{%
  \bibinfo {author} {\bibfnamefont{J.~R.}\ \bibnamefont{Rubbmark}}, \bibinfo
  {author} {\bibfnamefont{M.~M.}\ \bibnamefont{Kash}}, \bibinfo {author}
  {\bibfnamefont{M.~G.}\ \bibnamefont{Littman}},\ and\ \bibinfo {author}
  {\bibfnamefont{D.}~\bibnamefont{Kleppner}},\ }%
  \bibfield{journal}{%
  \bibinfo {journal} {Phys. Rev. A}\ }%
  \textbf{\bibinfo {volume} {23}},\ \bibinfo {pages} {3107} (\bibinfo {year}
  {1981})%
  \bibAnnoteFile{NoStop}{PhysRevA.23.3107}%
\bibitem{Petta2010}%
  \BibitemOpen
  \bibfield{author}{%
  \bibinfo {author} {\bibfnamefont{J.~R.}\ \bibnamefont{Petta}}, \bibinfo
  {author} {\bibfnamefont{H.}~\bibnamefont{Lu}},\ and\ \bibinfo {author}
  {\bibfnamefont{A.~C.}\ \bibnamefont{Gossard}},\ }%
  \bibfield{journal}{%
  \bibinfo {journal} {Science}\ }%
  \textbf{\bibinfo {volume} {327}},\ \bibinfo {pages} {669} (\bibinfo {year}
  {2010})%
  \bibAnnoteFile{NoStop}{Petta2010}%
\bibitem{PhysRevLett.96.187002}%
  \BibitemOpen
  \bibfield{author}{%
  \bibinfo {author} {\bibfnamefont{M.}~\bibnamefont{Sillanp\"a\"a}}, \bibinfo
  {author} {\bibfnamefont{T.}~\bibnamefont{Lehtinen}}, \bibinfo {author}
  {\bibfnamefont{A.}~\bibnamefont{Paila}}, \bibinfo {author}
  {\bibfnamefont{Y.}~\bibnamefont{Makhlin}},\ and\ \bibinfo {author}
  {\bibfnamefont{P.}~\bibnamefont{Hakonen}},\ }%
  \bibfield{journal}{%
  \bibinfo {journal} {Phys. Rev. Lett.}\ }%
  \textbf{\bibinfo {volume} {96}},\ \bibinfo {pages} {187002} (\bibinfo {year}
  {2006})%
  \bibAnnoteFile{NoStop}{PhysRevLett.96.187002}%
\bibitem{Fuchs2011a}%
  \BibitemOpen
  \bibfield{author}{%
  \bibinfo {author} {\bibfnamefont{G.~D.}\ \bibnamefont{Fuchs}}, \bibinfo
  {author} {\bibfnamefont{G.}~\bibnamefont{Burkard}}, \bibinfo {author}
  {\bibfnamefont{P.~V.}\ \bibnamefont{Klimov}},\ and\ \bibinfo {author}
  {\bibfnamefont{D.~D.}\ \bibnamefont{Awschalom}},\ }%
  \bibfield{journal}{%
  \bibinfo {journal} {Nat Phys}\ }%
  \textbf{\bibinfo {volume} {7}},\ \bibinfo {pages} {789} (\bibinfo {year}
  {2011})%
  \bibAnnoteFile{NoStop}{Fuchs2011a}%
\bibitem{1990PhRvL..65.2642S}%
  \BibitemOpen
  \bibfield{author}{%
  \bibinfo {author} {\bibfnamefont{R.~J.~C.}\ \bibnamefont{{Spreeuw}}},
  \bibinfo {author} {\bibfnamefont{N.~J.}\ \bibnamefont{{van Druten}}},
  \bibinfo {author} {\bibfnamefont{M.~W.}\ \bibnamefont{{Beijersbergen}}},
  \bibinfo {author} {\bibfnamefont{E.~R.}\ \bibnamefont{{Eliel}}},\ and\
  \bibinfo {author} {\bibfnamefont{J.~P.}\ \bibnamefont{{Woerdman}}},\ }%
  \bibfield{journal}{%
  \bibinfo {journal} {Physical Review Letters}\ }%
  \textbf{\bibinfo {volume} {65}},\ \bibinfo {pages} {2642} (\bibinfo {year}
  {1990})%
  \bibAnnoteFile{NoStop}{1990PhRvL..65.2642S}%
\bibitem{1995PhRvA..51..646B}%
  \BibitemOpen
  \bibfield{author}{%
  \bibinfo {author} {\bibfnamefont{D.}~\bibnamefont{{Bouwmeester}}}, \bibinfo
  {author} {\bibfnamefont{N.~H.}\ \bibnamefont{{Dekker}}}, \bibinfo {author}
  {\bibfnamefont{F.~E.~V.}\ \bibnamefont{{Dorsselaer}}}, \bibinfo {author}
  {\bibfnamefont{C.~A.}\ \bibnamefont{{Schrama}}}, \bibinfo {author}
  {\bibfnamefont{P.~M.}\ \bibnamefont{{Visser}}},\ and\ \bibinfo {author}
  {\bibfnamefont{J.~P.}\ \bibnamefont{{Woerdman}}},\ }%
  \bibfield{journal}{%
  \bibinfo {journal} {\pra}\ }%
  \textbf{\bibinfo {volume} {51}},\ \bibinfo {pages} {646} (\bibinfo {year}
  {1995})%
  \bibAnnoteFile{NoStop}{1995PhRvA..51..646B}%
\bibitem{PhysRevE.51.1861}%
  \BibitemOpen
  \bibfield{author}{%
  \bibinfo {author} {\bibfnamefont{J.}~\bibnamefont{Koz\l{}owski}}, \bibinfo
  {author} {\bibfnamefont{U.}~\bibnamefont{Parlitz}},\ and\ \bibinfo {author}
  {\bibfnamefont{W.}~\bibnamefont{Lauterborn}},\ }%
  \bibfield{journal}{%
  \Doi{10.1103/PhysRevE.51.1861}{\bibinfo {journal} {Phys. Rev. E}}\ }%
  \textbf{\bibinfo {volume} {51}},\ \bibinfo {pages} {1861} (\bibinfo {year}
  {1995})%
  \bibAnnoteFile{NoStop}{PhysRevE.51.1861}%
\bibitem{verbridge:124304}%
  \BibitemOpen
  \bibfield{author}{%
  \bibinfo {author} {\bibfnamefont{S.~S.}\ \bibnamefont{Verbridge}}, \bibinfo
  {author} {\bibfnamefont{J.~M.}\ \bibnamefont{Parpia}}, \bibinfo {author}
  {\bibfnamefont{R.~B.}\ \bibnamefont{Reichenbach}}, \bibinfo {author}
  {\bibfnamefont{L.~M.}\ \bibnamefont{Bellan}},\ and\ \bibinfo {author}
  {\bibfnamefont{H.~G.}\ \bibnamefont{Craighead}},\ }%
  \bibfield{journal}{%
  \bibinfo {journal} {Journal of Applied Physics}\ }%
  \textbf{\bibinfo {volume} {99}},\ \bibinfo {eid} {124304} (\bibinfo {year}
  {2006})%
  \bibAnnoteFile{NoStop}{verbridge:124304}%
\bibitem{PhysRevLett.105.027205}%
  \BibitemOpen
  \bibfield{author}{%
  \bibinfo {author} {\bibfnamefont{Q.~P.}\ \bibnamefont{Unterreithmeier}},
  \bibinfo {author} {\bibfnamefont{T.}~\bibnamefont{Faust}},\ and\ \bibinfo
  {author} {\bibfnamefont{J.~P.}\ \bibnamefont{Kotthaus}},\ }%
  \bibfield{journal}{%
  \bibinfo {journal} {Phys. Rev. Lett.}\ }%
  \textbf{\bibinfo {volume} {105}},\ \bibinfo {pages} {027205} (\bibinfo {year}
  {2010})%
  \bibAnnoteFile{NoStop}{PhysRevLett.105.027205}%
\bibitem{Unterreithmeier2009}%
  \BibitemOpen
  \bibfield{author}{%
  \bibinfo {author} {\bibfnamefont{Q.~P.}\ \bibnamefont{Unterreithmeier}},
  \bibinfo {author} {\bibfnamefont{E.~M.}\ \bibnamefont{Weig}},\ and\ \bibinfo
  {author} {\bibfnamefont{J.~P.}\ \bibnamefont{Kotthaus}},\ }%
  \bibfield{journal}{%
  \bibinfo {journal} {Nature}\ }%
  \textbf{\bibinfo {volume} {458}},\ \bibinfo {pages} {1001} (\bibinfo {year}
  {2009})%
  \bibAnnoteFile{NoStop}{Unterreithmeier2009}%
\bibitem{Groblacher2009}%
  \BibitemOpen
  \bibfield{author}{%
  \bibinfo {author} {\bibfnamefont{S.}~\bibnamefont{Groblacher}}, \bibinfo
  {author} {\bibfnamefont{K.}~\bibnamefont{Hammerer}}, \bibinfo {author}
  {\bibfnamefont{M.~R.}\ \bibnamefont{Vanner}},\ and\ \bibinfo {author}
  {\bibfnamefont{M.}~\bibnamefont{Aspelmeyer}},\ }%
  \bibfield{journal}{%
  \bibinfo {journal} {Nature}\ }%
  \textbf{\bibinfo {volume} {460}},\ \bibinfo {pages} {724} (\bibinfo {year}
  {2009})%
  \bibAnnoteFile{NoStop}{Groblacher2009}%
\bibitem{2011arXiv1107.3761V}%
  \BibitemOpen
  \bibfield{author}{%
  \bibinfo {author} {\bibfnamefont{E.}~\bibnamefont{{Verhagen}}}, \bibinfo
  {author} {\bibfnamefont{S.}~\bibnamefont{{Del{\'e}glise}}}, \bibinfo {author}
  {\bibfnamefont{S.}~\bibnamefont{{Weis}}}, \bibinfo {author}
  {\bibfnamefont{A.}~\bibnamefont{{Schliesser}}},\ and\ \bibinfo {author}
  {\bibfnamefont{T.~J.}\ \bibnamefont{{Kippenberg}}},\ }%
  \bibfield{journal}{%
  \bibinfo {journal} {ArXiv e-prints}}%
   (\bibinfo {year} {2011}),\
  \Eprint{http://arxiv.org/abs/1107.3761}{arXiv:1107.3761 [quant-ph]}%
  \bibAnnoteFile{NoStop}{2011arXiv1107.3761V}%
\bibitem{Teufel2011}%
  \BibitemOpen
  \bibfield{author}{%
  \bibinfo {author} {\bibfnamefont{J.~D.}\ \bibnamefont{Teufel}}, \bibinfo
  {author} {\bibfnamefont{T.}~\bibnamefont{Donner}}, \bibinfo {author}
  {\bibfnamefont{D.}~\bibnamefont{Li}}, \bibinfo {author}
  {\bibfnamefont{J.~W.}\ \bibnamefont{Harlow}}, \bibinfo {author}
  {\bibfnamefont{M.~S.}\ \bibnamefont{Allman}}, \bibinfo {author}
  {\bibfnamefont{K.}~\bibnamefont{Cicak}}, \bibinfo {author}
  {\bibfnamefont{A.~J.}\ \bibnamefont{Sirois}}, \bibinfo {author}
  {\bibfnamefont{J.~D.}\ \bibnamefont{Whittaker}}, \bibinfo {author}
  {\bibfnamefont{K.~W.}\ \bibnamefont{Lehnert}},\ and\ \bibinfo {author}
  {\bibfnamefont{R.~W.}\ \bibnamefont{Simmonds}},\ }%
  \bibfield{journal}{%
  \bibinfo {journal} {Nature}\ }%
  \textbf{\bibinfo {volume} {475}},\ \bibinfo {pages} {359} (\bibinfo {year}
  {2011})%
  \bibAnnoteFile{NoStop}{Teufel2011}%
\bibitem{Chan2011}%
  \BibitemOpen
  \bibfield{author}{%
  \bibinfo {author} {\bibfnamefont{J.}~\bibnamefont{Chan}}, \bibinfo {author}
  {\bibfnamefont{T.~P.~M.}\ \bibnamefont{Alegre}}, \bibinfo {author}
  {\bibfnamefont{A.~H.}\ \bibnamefont{Safavi-Naeini}}, \bibinfo {author}
  {\bibfnamefont{J.~T.}\ \bibnamefont{Hill}}, \bibinfo {author}
  {\bibfnamefont{A.}~\bibnamefont{Krause}}, \bibinfo {author}
  {\bibfnamefont{S.}~\bibnamefont{Groblacher}}, \bibinfo {author}
  {\bibfnamefont{M.}~\bibnamefont{Aspelmeyer}},\ and\ \bibinfo {author}
  {\bibfnamefont{O.}~\bibnamefont{Painter}},\ }%
  \bibfield{journal}{%
  \bibinfo {journal} {Nature}\ }%
  \textbf{\bibinfo {volume} {478}},\ \bibinfo {pages} {89} (\bibinfo {year}
  {2011})%
  \bibAnnoteFile{NoStop}{Chan2011}%
\bibitem{OConnell2010}%
  \BibitemOpen
  \bibfield{author}{%
  \bibinfo {author} {\bibfnamefont{A.~D.}\ \bibnamefont{O'Connell}}, \bibinfo
  {author} {\bibfnamefont{M.}~\bibnamefont{Hofheinz}}, \bibinfo {author}
  {\bibfnamefont{M.}~\bibnamefont{Ansmann}}, \bibinfo {author}
  {\bibfnamefont{R.~C.}\ \bibnamefont{Bialczak}}, \bibinfo {author}
  {\bibfnamefont{M.}~\bibnamefont{Lenander}}, \bibinfo {author}
  {\bibfnamefont{E.}~\bibnamefont{Lucero}}, \bibinfo {author}
  {\bibfnamefont{M.}~\bibnamefont{Neeley}}, \bibinfo {author}
  {\bibfnamefont{D.}~\bibnamefont{Sank}}, \bibinfo {author}
  {\bibfnamefont{H.}~\bibnamefont{Wang}}, \bibinfo {author}
  {\bibfnamefont{M.}~\bibnamefont{Weides}}, \bibinfo {author}
  {\bibfnamefont{J.}~\bibnamefont{Wenner}}, \bibinfo {author}
  {\bibfnamefont{J.~M.}\ \bibnamefont{Martinis}},\ and\ \bibinfo {author}
  {\bibfnamefont{A.~N.}\ \bibnamefont{Cleland}},\ }%
  \bibfield{journal}{%
  \bibinfo {journal} {Nature}\ }%
  \textbf{\bibinfo {volume} {464}},\ \bibinfo {pages} {697} (\bibinfo {year}
  {2010})%
  \bibAnnoteFile{NoStop}{OConnell2010}%
\bibitem{APEX.2.062202}%
  \BibitemOpen
  \bibfield{author}{%
  \bibinfo {author} {\bibfnamefont{H.}~\bibnamefont{Okamoto}}, \bibinfo
  {author} {\bibfnamefont{T.}~\bibnamefont{Kamada}}, \bibinfo {author}
  {\bibfnamefont{K.}~\bibnamefont{Onomitsu}}, \bibinfo {author}
  {\bibfnamefont{I.}~\bibnamefont{Mahboob}},\ and\ \bibinfo {author}
  {\bibfnamefont{H.}~\bibnamefont{Yamaguchi}},\ }%
  \bibfield{journal}{%
  \bibinfo {journal} {Applied Physics Express}\ }%
  \textbf{\bibinfo {volume} {2}},\ \bibinfo {pages} {062202} (\bibinfo {year}
  {2009})%
  \bibAnnoteFile{NoStop}{APEX.2.062202}%
\bibitem{PhysRevB.79.165309}%
  \BibitemOpen
  \bibfield{author}{%
  \bibinfo {author} {\bibfnamefont{R.~B.}\ \bibnamefont{Karabalin}}, \bibinfo
  {author} {\bibfnamefont{M.~C.}\ \bibnamefont{Cross}},\ and\ \bibinfo {author}
  {\bibfnamefont{M.~L.}\ \bibnamefont{Roukes}},\ }%
  \bibfield{journal}{%
  \bibinfo {journal} {Phys. Rev. B}\ }%
  \textbf{\bibinfo {volume} {79}},\ \bibinfo {pages} {165309} (\bibinfo {year}
  {2009})%
  \bibAnnoteFile{NoStop}{PhysRevB.79.165309}%
\bibitem{PhysRevLett.106.094102}%
  \BibitemOpen
  \bibfield{author}{%
  \bibinfo {author} {\bibfnamefont{R.~B.}\ \bibnamefont{Karabalin}}, \bibinfo
  {author} {\bibfnamefont{R.}~\bibnamefont{Lifshitz}}, \bibinfo {author}
  {\bibfnamefont{M.~C.}\ \bibnamefont{Cross}}, \bibinfo {author}
  {\bibfnamefont{M.~H.}\ \bibnamefont{Matheny}}, \bibinfo {author}
  {\bibfnamefont{S.~C.}\ \bibnamefont{Masmanidis}},\ and\ \bibinfo {author}
  {\bibfnamefont{M.~L.}\ \bibnamefont{Roukes}},\ }%
  \bibfield{journal}{%
  \bibinfo {journal} {Phys. Rev. Lett.}\ }%
  \textbf{\bibinfo {volume} {106}},\ \bibinfo {pages} {094102} (\bibinfo {year}
  {2011})%
  \bibAnnoteFile{NoStop}{PhysRevLett.106.094102}%
\bibitem{perisanu:063110}%
  \BibitemOpen
  \bibfield{author}{%
  \bibinfo {author} {\bibfnamefont{S.}~\bibnamefont{Perisanu}}, \bibinfo
  {author} {\bibfnamefont{T.}~\bibnamefont{Barois}}, \bibinfo {author}
  {\bibfnamefont{P.}~\bibnamefont{Poncharal}}, \bibinfo {author}
  {\bibfnamefont{T.}~\bibnamefont{Gaillard}}, \bibinfo {author}
  {\bibfnamefont{A.}~\bibnamefont{Ayari}}, \bibinfo {author}
  {\bibfnamefont{S.~T.}\ \bibnamefont{Purcell}},\ and\ \bibinfo {author}
  {\bibfnamefont{P.}~\bibnamefont{Vincent}},\ }%
  \bibfield{journal}{%
  \bibinfo {journal} {Applied Physics Letters}\ }%
  \textbf{\bibinfo {volume} {98}},\ \bibinfo {eid} {063110} (\bibinfo {year}
  {2011})%
  \bibAnnoteFile{NoStop}{perisanu:063110}%
\bibitem{PhysRevLett.105.117205}%
  \BibitemOpen
  \bibfield{author}{%
  \bibinfo {author} {\bibfnamefont{H.~J.~R.}\ \bibnamefont{Westra}}, \bibinfo
  {author} {\bibfnamefont{M.}~\bibnamefont{Poot}}, \bibinfo {author}
  {\bibfnamefont{H.~S.~J.}\ \bibnamefont{van~der Zant}},\ and\ \bibinfo
  {author} {\bibfnamefont{W.~J.}\ \bibnamefont{Venstra}},\ }%
  \bibfield{journal}{%
  \bibinfo {journal} {Phys. Rev. Lett.}\ }%
  \textbf{\bibinfo {volume} {105}},\ \bibinfo {pages} {117205} (\bibinfo {year}
  {2010})%
  \bibAnnoteFile{NoStop}{PhysRevLett.105.117205}%
\bibitem{kozinsky:253101}%
  \BibitemOpen
  \bibfield{author}{%
  \bibinfo {author} {\bibfnamefont{I.}~\bibnamefont{Kozinsky}}, \bibinfo
  {author} {\bibfnamefont{H.~W.~C.}\ \bibnamefont{Postma}}, \bibinfo {author}
  {\bibfnamefont{I.}~\bibnamefont{Bargatin}},\ and\ \bibinfo {author}
  {\bibfnamefont{M.~L.}\ \bibnamefont{Roukes}},\ }%
  \bibfield{journal}{%
  \Doi{10.1063/1.2209211}{\bibinfo {journal} {Applied Physics Letters}}\ }%
  \textbf{\bibinfo {volume} {88}},\ \bibinfo {eid} {253101} (\bibinfo {year}
  {2006})%
  \bibAnnoteFile{NoStop}{kozinsky:253101}%
\bibitem{2011arXiv1109.1156F}%
  \BibitemOpen
  \bibfield{author}{%
  \bibinfo {author} {\bibfnamefont{T.}~\bibnamefont{{Faust}}}, \bibinfo
  {author} {\bibfnamefont{P.}~\bibnamefont{{Krenn}}}, \bibinfo {author}
  {\bibfnamefont{S.}~\bibnamefont{{Manus}}}, \bibinfo {author}
  {\bibfnamefont{J.~P.}\ \bibnamefont{{Kotthaus}}},\ and\ \bibinfo {author}
  {\bibfnamefont{E.~M.}\ \bibnamefont{{Weig}}},\ }%
  \bibfield{journal}{%
  \Doi{10.1038/ncomms1723}{\bibinfo {journal} {Nature Communications}}\ }%
  \textbf{\bibinfo {volume} {3}},\ \bibinfo {pages} {728} (\bibinfo {year}
  {2012})%
  \bibAnnoteFile{NoStop}{2011arXiv1109.1156F}%
\bibitem{schmid:163506}%
  \BibitemOpen
  \bibfield{author}{%
  \bibinfo {author} {\bibfnamefont{S.}~\bibnamefont{Schmid}}, \bibinfo {author}
  {\bibfnamefont{M.}~\bibnamefont{Wendlandt}}, \bibinfo {author}
  {\bibfnamefont{D.}~\bibnamefont{Junker}},\ and\ \bibinfo {author}
  {\bibfnamefont{C.}~\bibnamefont{Hierold}},\ }%
  \bibfield{journal}{%
  \Doi{10.1063/1.2362590}{\bibinfo {journal} {Applied Physics Letters}}\ }%
  \textbf{\bibinfo {volume} {89}},\ \bibinfo {eid} {163506} (\bibinfo {year}
  {2006})%
  \bibAnnoteFile{NoStop}{schmid:163506}%
\bibitem{APL}%
  \BibitemOpen
  \bibfield{author}{%
  \bibinfo {author} {\bibfnamefont{J.}~\bibnamefont{Rieger}}, \bibinfo {author}
  {\bibfnamefont{T.}~\bibnamefont{{Faust}}}, \bibinfo {author}
  {\bibfnamefont{M.~J.}\ \bibnamefont{Seitner}}, \bibinfo {author}
  {\bibfnamefont{J.~P.}\ \bibnamefont{{Kotthaus}}},\ and\ \bibinfo {author}
  {\bibfnamefont{E.~M.}\ \bibnamefont{{Weig}}},\ }%
  \bibfield{journal}{%
  \bibinfo {journal} {{in preparation }}}%
   (\bibinfo {year} {2012})%
  \bibAnnoteFile{NoStop}{APL}%
\bibitem{supp}%
  \BibitemOpen
  \bibinfo {journal} {See Supplemental Material for the solution of the coupled
  differential equations, details of the tuning behaviour, the data analysis
  and the coupling mechanism}%
  \bibAnnoteFile{NoStop}{supp}%
\bibitem{10.1119/1.3471177}%
  \BibitemOpen
\bibfield{journal}{%
    }%
  \bibfield{author}{%
  \bibinfo {author} {\bibfnamefont{L.}~\bibnamefont{Novotny}},\ }%
  \bibfield{journal}{%
  \bibinfo {journal} {American Journal of Physics}\ }%
  \textbf{\bibinfo {volume} {78}},\ \bibinfo {pages} {1199} (\bibinfo {year}
  {2010})%
  \bibAnnoteFile{NoStop}{10.1119/1.3471177}%
\bibitem{PhysRevA.66.023404}%
  \BibitemOpen
  \bibfield{author}{%
  \bibinfo {author} {\bibfnamefont{J.}~\bibnamefont{Liu}}, \bibinfo {author}
  {\bibfnamefont{L.}~\bibnamefont{Fu}}, \bibinfo {author}
  {\bibfnamefont{B.-Y.}\ \bibnamefont{Ou}}, \bibinfo {author}
  {\bibfnamefont{S.-G.}\ \bibnamefont{Chen}}, \bibinfo {author}
  {\bibfnamefont{D.-I.}\ \bibnamefont{Choi}}, \bibinfo {author}
  {\bibfnamefont{B.}~\bibnamefont{Wu}},\ and\ \bibinfo {author}
  {\bibfnamefont{Q.}~\bibnamefont{Niu}},\ }%
  \bibfield{journal}{%
  \bibinfo {journal} {Phys. Rev. A}\ }%
  \textbf{\bibinfo {volume} {66}},\ \bibinfo {pages} {023404} (\bibinfo {year}
  {2002})%
  \bibAnnoteFile{NoStop}{PhysRevA.66.023404}%
\bibitem{1367-2630-10-7-073008}%
  \BibitemOpen
  \bibfield{author}{%
  \bibinfo {author} {\bibfnamefont{Q.}~\bibnamefont{{Zhang}}}, \bibinfo
  {author} {\bibfnamefont{P.}~\bibnamefont{{H{\"a}nggi}}},\ and\ \bibinfo
  {author} {\bibfnamefont{J.}~\bibnamefont{{Gong}}},\ }%
  \bibfield{journal}{%
  \bibinfo {journal} {New Journal of Physics}\ }%
  \textbf{\bibinfo {volume} {10}},\ \bibinfo {pages} {073008} (\bibinfo {year}
  {2008})%
  \bibAnnoteFile{NoStop}{1367-2630-10-7-073008}%
\bibitem{PhysRevLett.107.043603}%
  \BibitemOpen
  \bibfield{author}{%
  \bibinfo {author} {\bibfnamefont{G.}~\bibnamefont{Heinrich}}, \bibinfo
  {author} {\bibfnamefont{M.}~\bibnamefont{Ludwig}}, \bibinfo {author}
  {\bibfnamefont{J.}~\bibnamefont{Qian}}, \bibinfo {author}
  {\bibfnamefont{B.}~\bibnamefont{Kubala}},\ and\ \bibinfo {author}
  {\bibfnamefont{F.}~\bibnamefont{Marquardt}},\ }%
  \bibfield{journal}{%
  \Doi{10.1103/PhysRevLett.107.043603}{\bibinfo {journal} {Phys. Rev. Lett.}}\
  }%
  \textbf{\bibinfo {volume} {107}},\ \bibinfo {pages} {043603} (\bibinfo {year}
  {2011})%
  \bibAnnoteFile{NoStop}{PhysRevLett.107.043603}%
\bibitem{2011arXiv1111.2967L}%
  \BibitemOpen
  \bibfield{author}{%
  \bibinfo {author} {\bibfnamefont{R.}~\bibnamefont{{Lifshitz}}}, \bibinfo
  {author} {\bibfnamefont{E.}~\bibnamefont{{Kenig}}},\ and\ \bibinfo {author}
  {\bibfnamefont{M.~C.}\ \bibnamefont{{Cross}}},\ }%
  \bibfield{journal}{%
  \bibinfo {journal} {ArXiv e-prints}}%
   (\bibinfo {year} {2011}),\
  \Eprint{http://arxiv.org/abs/1111.2967}{arXiv:1111.2967 [nlin.PS]}%
  \bibAnnoteFile{NoStop}{2011arXiv1111.2967L}%
\end{thebibliography}%

\renewcommand{\thefigure}{S\arabic{figure}}
 \renewcommand{\theequation}{S\arabic{equation}}
  \setcounter{figure}{0}
\setcounter{equation}{0}
\renewcommand*{\theHfigure}{\thepart.\thefigure}
\renewcommand*{\theHequation}{\thepart.\theequation}
\newcommand{\eq}[1]{equation~(\ref{#1})}
\onecolumngrid
\clearpage

\section{Supplemental Material to "Non-adiabatic dynamics of two strongly coupled nanomechanical resonator modes"}

\pagestyle{empty}

\subsection{Modeling and data analysis}

\subsubsection{Modeling the anticrossing}

The two coupled differential equations for the displacements $u_n$ (n=1,2) are
\begin{eqnarray}
\label{dgl1}
m_{\rm eff}\frac{d^2u_1}{d t^2}+m_{\rm eff}\gamma_1\frac{d u_1}{d t} +k_1u_1+k_c(u_1-u_2)=0 \\
\label{dgl2}
m_{\rm eff}\frac{d^2u_2}{d t^2}+m_{\rm eff}\gamma_2\frac{d u_2}{d t} +k_2u_2+k_c(u_2-u_1)=0,
\end{eqnarray}
using the effective mass of the resonator (which, for a doubly clamped beam, is half its total mass) $m_{\rm eff}$, the spring constants $k_n$, the coupling constant $k_c$ and the two different damping constants $\gamma_n$. They are solved using an oscillatory ansatz $u_n=a_ne^{i\omega t}$. This leads to 
\begin{eqnarray}
-\omega^2m_{\rm eff}a_1+k_1a_1+k_c(a_1-a_2)+im_{\rm eff}\omega\gamma_1a_1=0 \\
-\omega^2m_{\rm eff}a_2+k_2a_2+k_c(a_2-a_1)+im_{\rm eff}\omega\gamma_2a_2=0
\end{eqnarray}
which can be rewritten as
\begin{eqnarray}
\left(\begin{array}{cc}
-\omega^2m_{\rm eff}+im_{\rm eff}\omega\gamma_1 & 0 \\
0 & -\omega^2m_{\rm eff}+im_{\rm eff}\omega\gamma_2
\end{array}\right)\vec{a}+
\left(\begin{array}{cc}
k_1+k_c & -k_c \\
-k_c & k_2+k_c
\end{array}\right)\vec{a} =0 \\
\boldsymbol\Omega \vec{a} + \boldsymbol K\vec{a}=0
\end{eqnarray}
using $\vec{a}=\left(\begin{array}{c}a_1 \\ a_2 \end{array}\right)$ and by defining the two matrices $\bf \Omega$ and $\bf K$.
By solving the eigenwert problem
\begin{equation}
\boldsymbol\Omega^{-1}\boldsymbol K\vec{a}=\lambda\vec{a}
\end{equation}
one can obtain the two (rather complicated) analytical solutions.
The real part of these solutions contains the frequencies while the imaginary part describes the damping constants of the two branches.

\subsubsection{Fitting the measured anticrossing}

\begin{figure}[ht]
\includegraphics{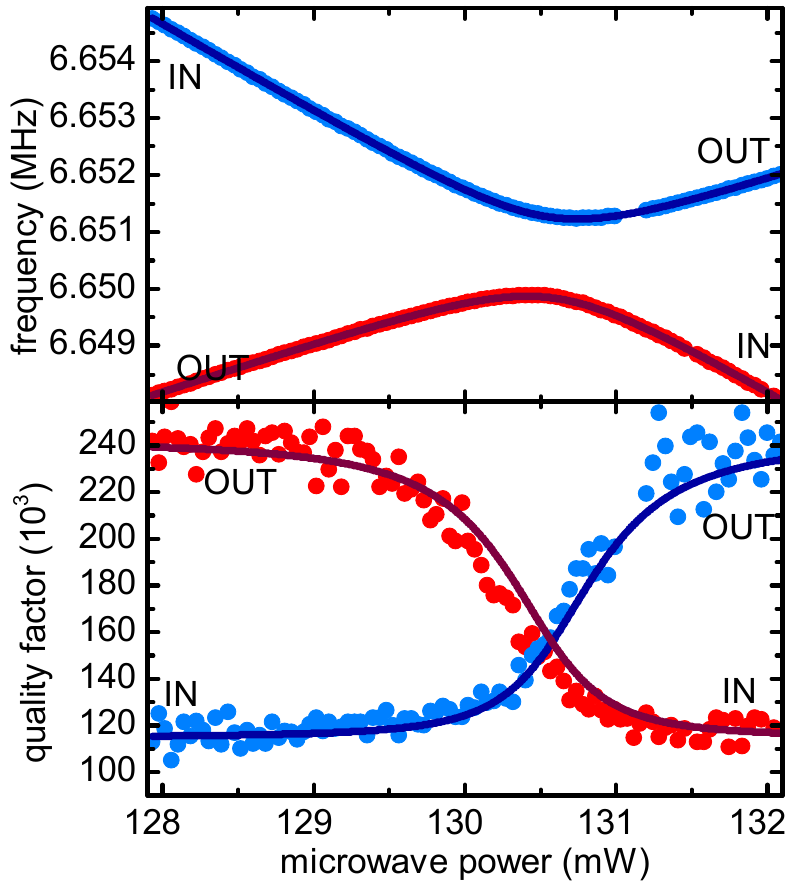}
\caption{\label{s1}Avoided crossing data and corresponding fit of a second resonator tuned via the microwave power, showing the coupling behavior of both frequency and quality factor. As the resonator modes are swept through the coupling region, the two branches transform from the high quality factor of the out of plane mode to the lower quality factor of the in plane mode and vice versa.}
\end{figure}

These two solutions are then simultaneously fitted to the measured frequencies and Q factors.
We approximate $k_n=k_0+\kappa_n(U-U_0)+\lambda_n(U-U_0)^2$.
This is a second order series expansion of the parabolic frequency tuning\,\cite{Unterreithmeier2009,2011arXiv1109.1156F} around the crossing voltage $U_0$.
The fit shown in Fig. 3 yields the following values: $k_0=3.2\,\frac{\rm N}{\rm m}$, $\kappa_1=9.3\cdot10^{-3}\,\frac{\rm N}{\rm Vm}, \kappa_2=-12.5\cdot10^{-3}\,\frac{\rm N}{\rm Vm}, \lambda_1=0.41\cdot10^{-3}\,\frac{\rm N}{\rm V^2m}, \lambda_2=-0.57\cdot10^{-3}\,\frac{\rm N}{\rm V^2m}, k_c=3.76\cdot10^{-3}\,\frac{\rm N}{\rm m}$ and $U_0=6.547\,\rm V$ using an effective mass of $m_{\rm eff}=1.85\cdot10^{-15}$g.
As $|U-U_0|$ is always less than $3\,\rm V$ in the experiment, the maximal relative influence of the quadratic term $\frac{|U-U_0|\lambda_n}{\kappa_n}$ is below 15\,\%.

The fit in Fig.~3 shows only the measured frequencies, as the quality factors of the two modes are nearly identical.
A resonator on a different chip which was tuned using the microwave power instead of the DC voltage\,\cite{2011arXiv1109.1156F}, exhibited higher, dissimmilar, quality factors and thus allowed to simultaneously fit frequencies and quality factors.
The resulting graphs are shown in \fig{s1}.
One can clearly see how the quality factors of the red and blue branch change as the system is tuned through the coupling region and the oscillation transforms from an in plane to an out of plane motion and vice versa.

\subsubsection{Analyzing the time-resolved data displayed in Fig.~4}

For each ramp time $\tau$, we measure the time-dependent power spectral density at points A  and D of Fig.~2 with a bandwidth of $1\,$kHz to have sufficient time resolution.
The measurement is started at $t=\tau$, but the data used for the analysis is the one taken after time $\delta=3\,$ms to avoid transient spikes in the measurement (as illustrated in Fig.~4a).
The resulting exponential decays are then fitted using $S^{A,D}(t,\tau)=S_0^{A,D}(\tau)e^{-\gamma t}+S_{\rm Noise}$, yielding the noise floor $S_{\rm Noise}$, the damping constant $\gamma$, identical to the one determined from spectral measurements, and the mode energy at time $\tau$ $S^{A,D}_0(\tau)$ at points A and D.

\subsubsection{Converting the measured data into transition probabilities}

The transition probabilities of a classical Landau-Zener-Transition\,\cite{10.1119/1.3471177} are
\begin{equation}
P_{\rm dia}=e^{\frac{-\pi\Gamma^2}{2\alpha}}\ \rm{and}\ P_{\rm adia}=1-P_{\rm dia}.
\end{equation}
By rewriting the change of the frequency difference between the two unperturbed states $\omega_n(U)=\sqrt{\frac{k_0+k_n(U)}{m_{\rm eff}}}$
\begin{equation}
\alpha=\frac{\partial (\omega_1-\omega_2)}{\partial t}=\frac{(\omega_1(U_i)-\omega_2(U_i))-(\omega_1(U_f)-\omega_2(U_f))}{\tau}=\frac{\Delta\omega}{\tau},
\end{equation}
$P_{\rm diab}$ can be expressed as a function of the ramp time $\tau$ and the measured frequency differences $\Delta\omega$ between the initial voltage $U_i$ and the final voltage $U_f$, as the two voltages are kept constant throughout the experiment and only $\tau$ is varied.
By introducing the mechanical damping term $e^{-\gamma t}$ at $t=\tau$, the probability is transformed into a normalized state population
\begin{eqnarray}
\label{sadia}
S_{\rm dia}(\tau)=e^{\frac{-\pi\Gamma^2\tau}{2\Delta\omega}-\gamma\tau} \\
S_{\rm adia}(\tau)=\left(1-e^{\frac{-\pi\Gamma^2\tau}{2\Delta\omega}}\right)e^{-\gamma\tau}.
\end{eqnarray}
The amplitudes of the two datasets $S^{A}_0(\tau)$ and $S^{D}_0(\tau)$ are then rescaled to fit these two equations. Both are shown in Fig.~4b of the main text along with the theoretical curves given in (\ref{sadia}).
Note that $\Gamma$, $\gamma$ and $\Delta\omega$ are already known from previous measurements and are no fit parameters.

\subsection{Coupling mechanism}

\subsubsection{Hybrid mode shapes}

\begin{figure}[ht]
\includegraphics{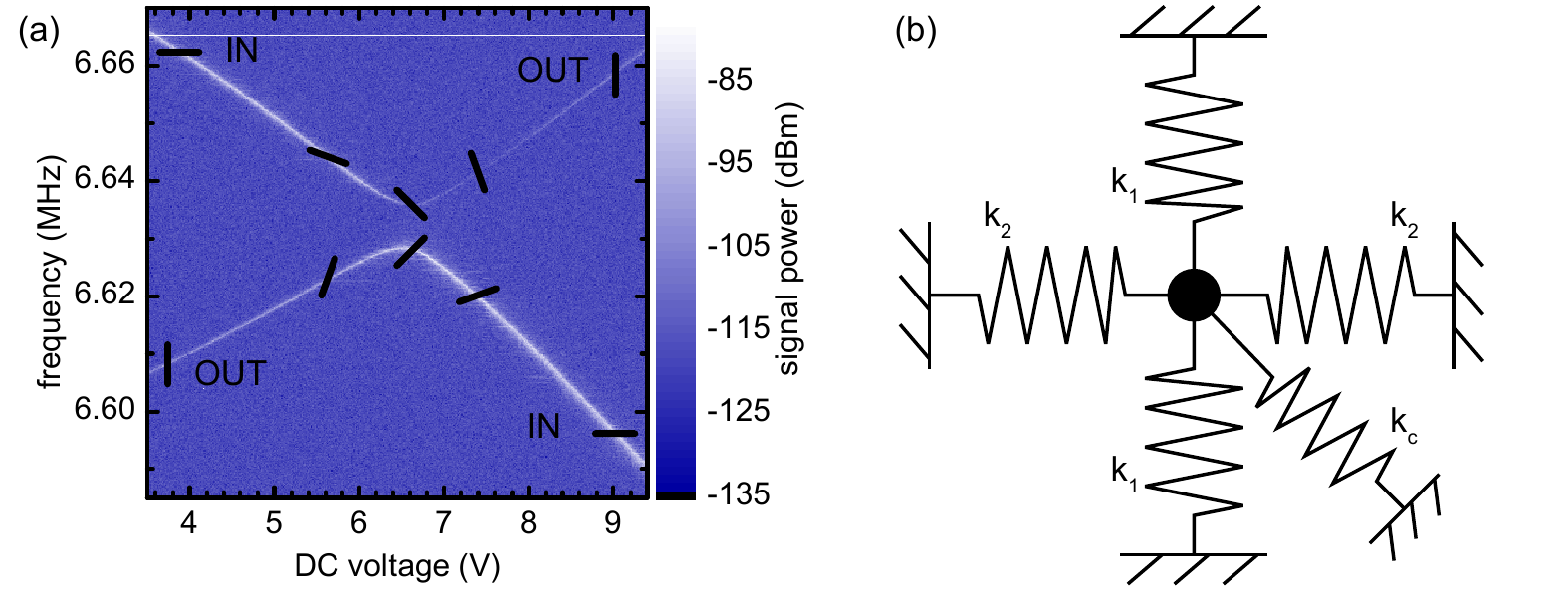}
\caption{\label{s2} The polarizations of the hybrid modes in the coupling region are sketched as black lines in the anticrossing diagram shown in (a). A horizontal line represents the in-plane and a vertical line represents the out-of-plane mode. The two hybrid modes exhibit polarization directions rotating in the plane perpendicular to the resonator.
A mass on a spring model of the resonator with an asymmetric ``coupling spring'' is displayed in (b).}
\end{figure}

To learn more about the coupling and to understand how the transformation from an in-plane to an out-of-plane motion (or vice versa) takes place during an adiabatic transition, it is interesting to look at the spatial mode profiles in the anticrossing region.
The solutions of the differential equations \ref{dgl1} and \ref{dgl2} are an in phase ($\omega=\sqrt{\frac{k}{m_{\rm eff}}}$) and an out of phase ($\omega=\sqrt{\frac{k+2k_c}{m_{\rm eff}}}$) combination of the fundamental mechanical modes.
The mode polarizations and their qualitative evolution throughout the coupling region is sketched in \fig[a]{s2}, showing the transition between the pure in- and out-of-plane modes via the diagonal hybrid modes.

These two diagonal hybrid modes have different frequencies and thus different energies.
In a perfectly symmetric beam with a rectangular cross section one would not expect any difference between the two diagonal modes, thus the coupling between the modes has to be connected to some asymmetry.
This is visualized in \fig[b]{s2}: the two springs labeled $k_1$ provide the restoring force of the out-of-plane mode, the springs $k_2$ correspond to the in-plane mode.
One can directly see from the schematic that the coupling spring $k_c$ introduces an asymmetry into the system.
One coupling mechanism, related to the asymmetric beam position between the electrodes, is discussed in the next section.

\subsubsection{Electrical field coupling}

\begin{figure}[ht]
\includegraphics{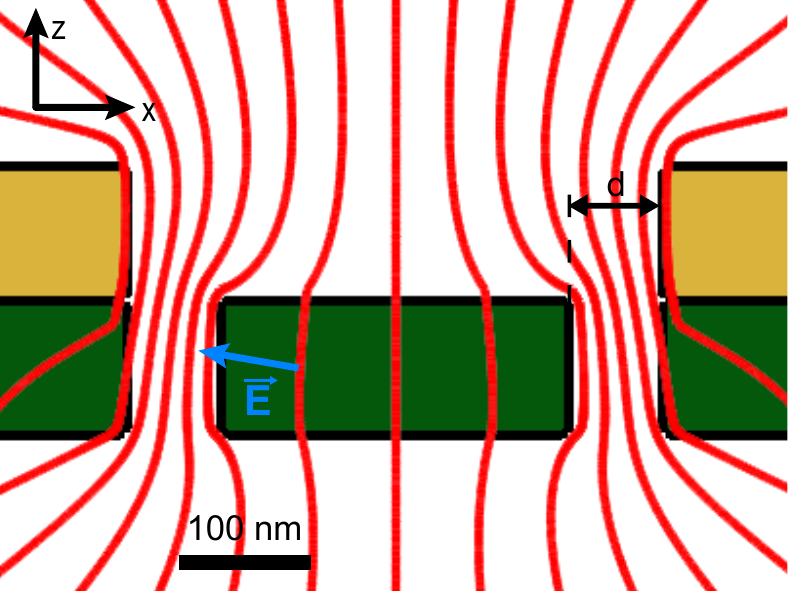}
\caption{\label{s3} The red lines of equal electric potential between the two electrodes (calculated using COMSOL finite element simulation) demonstrate the inhomogenous electric field (blue arrow) in x as well as z direction.}
\end{figure}

A mechanical resonator chip with different gaps between the side electrodes and each beam exhibits different coupling constants $\Gamma$ for each resonator.
The smallest gap (of about 60 to 70\,nm) yields $\Gamma=7.77$\,kHz (as presented in the main paper).
Fitting the frequencies of the other beams in their respective coupling region gives a $\Gamma$ of $7.27\pm0.09\,$kHz, $6.10\pm0.02\,$kHz and $5.31\pm0.03\,$kHz with increasing gap size up to roughly 150\,nm.
As the electric field between the electrodes decreases with their increasing separation d (and the voltage is approximately constant), the coupling seems to be mediated by the electric field, even though the exact gap sizes are unknown and thus a quantitative relation can not be established.

This electrical coupling between the in-plane mode (oscillating in x direction) and the out-of-plane mode (oscillating in z direction) can also be shown using the following simple model:\\
Starting from one undamped coupled equation (a simpler version of \eq{dgl1})
\begin{equation}
F_x=k_xx+k_c(x-z),
\end{equation}
the coupling constant is just the derivative of $F_x$ in z-direction:
\begin{equation}
\frac{\partial F_x}{\partial z} = -k_c
\end{equation}
The electric force on the dielectric beam is the gradient of its energy $W$ in an external electric field $\vec{E}$
\begin{equation}
\vec{F_{\rm el}}=-\vec{\nabla} W=-\vec{\nabla}(\vec{p}\cdot\vec{E})=-\vec{\nabla}(\alpha E^2)
\end{equation}
using the polarizability $\alpha$.
Thus, the derivative of the x component of $\vec{F_{\rm el}}$ in z direction yields a dielectric coupling term
\begin{equation}
\frac{\partial}{\partial z}\left(-\frac{\partial \alpha E^2}{\partial x}\right)=-\alpha\frac{\partial^2 E^2}{\partial z \partial x}=-k_{c,el}
\end{equation}
As there is a gradient of the electric field in z direction (the electrodes are above the beam) and in x direction (from asymmetry, otherwise the in-plane mode would not tune with the applied DC voltage), $k_{c,el}$ is not zero and contributes at least partially of the observed coupling strength.
This is visualized in \fig{s3}: if the beam is not perfectly aligned between the two electrodes, the resulting effective electric field exhibits a gradient in x and z direction.
The field-dependent coupling mechanism also explains why the data shown in \fig{s1}, measured with a DC voltage of 0\,V, exhibits a weaker coupling of less than 2\,kHz (the microwave field used to detect the beam motion also leads to an electric field, but the effective voltage is smaller).
As the two modes can not be tuned into resonance without applying a (DC or microwave) electric field, it is not possible to test if there is also any purely mechanical coupling, e. g. caused by interactions between the modes mediated by the prestress of the beam or coupling effects in the shared clamping points of the two modes.

%\bibliography{mode_coupling}

\end{document}